%% file: main.tex
\documentclass[tinyml]{acmart}

\usepackage{amsmath,amsfonts} 
\usepackage[ruled,vlined]{algorithm2e}
\usepackage{graphicx}
\usepackage{textcomp}
\usepackage{xcolor}
\usepackage{tabularx}
\usepackage{caption}
\usepackage{subcaption}
\usepackage[inline]{enumitem}
\usepackage{bm}
\usepackage{bbm}
\usepackage{url}
\usepackage{amsthm}
\usepackage{wrapfig}
\usepackage{multicol}
\usepackage{multirow}

\AtBeginDocument{%
  \providecommand\BibTeX{{%
    \normalfont B\kern-0.5em{\scshape i\kern-0.25em b}\kern-0.8em\TeX}}}

\setcopyright{rightsretained}
\copyrightyear{2024}
\acmYear{2024}

\acmConference[tinyML'24]{tinyML Research Symposium}{April, 2024}{Burlingame, CA}

\begin{document}

\title{MicroHD: An Accuracy-Driven Optimization of Hyperdimensional Computing Algorithms for TinyML systems}



\author{Flavio Ponzina}
\affiliation{%
  \institution{University of California San Diego}
  \city{La Jolla}
  \country{California, USA}}

\author{Tajana Rosing}
\affiliation{%
  \institution{University of California San Diego}
  \city{La Jolla}
  \country{California, USA}}

\renewcommand{\shortauthors}{Flavio Ponzina and Tajana Rosing}

\begin{abstract}
  Hyperdimensional computing (HDC) is emerging as a promising AI approach that can effectively target TinyML applications thanks to its lightweight computing and memory requirements. Previous works on HDC showed that limiting the standard 10k dimensions of the hyperdimensional space to much lower values is possible, reducing even more HDC resource requirements. Similarly, other studies demonstrated that binary values can be used as elements of the generated hypervectors, leading to significant efficiency gains at the cost of some degree of accuracy degradation. Nevertheless, current optimization attempts do not concurrently co-optimize HDC hyper-parameters, and accuracy degradation is not directly controlled, resulting in sub-optimal HDC models providing several applications with unacceptable output qualities. 
  
  In this work, we propose \emph{MicroHD}, a novel accuracy-driven HDC optimization approach that iteratively tunes HDC hyper-parameters, reducing memory and computing requirements while ensuring user-defined accuracy levels. The proposed method can be applied to HDC implementations using different encoding functions, demonstrates good scalability for larger HDC workloads, and achieves compression and efficiency gains up to 200$\times$ when compared to baseline implementations for accuracy degradations lower than 1\%. 

\end{abstract}

\keywords{TinyML, edge AI, hyperdimensional computing, AI optimization}

\maketitle

\input{Introduction.tex}
\input{Background.tex}
\input{RelatedWorks.tex}
\input{Methodology.tex}
\input{ExpSetup.tex}

\input{ExpResults.tex}

\input{Conclusions.tex}
\input{acknowledgements.tex}

\bibliographystyle{ACM-Reference-Format}
\bibliography{biblio}

\end{document}

%% file: Introduction.tex
\section{Introduction}

The edge computing approach is becoming highly attractive in the area of TinyML as technological advancements enrich embedded systems with smaller and more energy-efficient resources. Processing the collected data at the edge offers several advantages compared to the standard cloud computing alternative, including lower amounts of data transfers, high responsiveness, and high privacy.
In the last years, we have witnessed an increasing deployment of artificial intelligence (AI) workloads in embedded systems, mainly because of the high output quality that AI algorithms offer in most industrial and scientific applications~\cite{sun2019ai, denby2019orbital}. Nevertheless, new challenges are rising as new AI models are growing in complexity to improve their performance, thus limiting the pool of TinyML applications that can benefit from AI. Indeed, recent machine learning and deep learning models such as convolutional neural networks (CNNs) or transformers are compute- and memory-intense and struggle to perform real-time applications while fitting the limited memory resources of edge devices. This becomes even more critical when considering continual learning~\cite{de2021continual} or federated learning~\cite{zhang2021survey} settings, where on-device training, typically more complex than simple inference, is required~\cite{lin2022device}.

Hyperdimensional computing (HDC) aims to solve this challenge by employing a lightweight brain-inspired computing approach~\cite{kanerva2009hyperdimensional}.
HDC encodes input data in a high-dimensional holistic space where only simple element-wise operations are needed to perform both inference and training. A limited number of model parameters, low computing requirements, and a high degree of computing parallelism are all factors that make HDC a promising candidate for TinyML. Moreover, HDC already demonstrated encouraging results in a plethora of different applications, ranging from activity recognition~\cite{yao2021radar} to personalized health monitoring~\cite{asgarinejad2020detection}.

Despite being way more efficient than CNNs, HDC still benefits from model optimizations that reduce memory and computing requirements to further improve energy efficiency in edge devices. Examples of HDC optimizations investigated in previous works include binarization/ternarization of the encoded hypervectors (HVs)~\cite{imani2019quanthd} and dimensionality reduction~\cite{hernandez2021onlinehd}. However, these methods are often applied empirically and exclusively, and, from memory, computing, and accuracy perspectives, optimal HDC settings are not properly determined. Interestingly, similar HDC optimizations can have a highly different impact on accuracy when applied to models targeting different tasks or facing different dataset complexity. 
Indeed, accuracy is a key element to consider when optimizing HDC workloads, as significant accuracy drops may prevent a real-life deployment of these models in several applications. As a consequence, empirical solutions proposed in previous works can either require long design explorations or return unacceptable low-accuracy models for TinyML applications.

In this paper, we address this challenge by introducing \emph{MicroHD}, a novel \emph{accuracy-driven} optimization of HDC algorithms, reducing memory and compute requirements while ensuring user-defined accuracy constraints are met. We consider HDC workloads employing two common encoding strategies and achieve resource requirement reductions of up to 266$\times$ when compared to baseline HDC implementations for limited accuracy loss. We also compare our results with previous works and observe that \emph{MicroHD} outperforms previous approaches by achieving 8$\times$ compression gains for less than 0.5\% accuracy drop with respect to state-of-the-art implementations. To the best of our knowledge, this is the first work proposing an accuracy-driven model compression in the realm of HDC for TinyML. 
Our contributions can be summarized as follows:

\begin{itemize}
    \item We analyze HDC workloads from memory and computing perspectives and propose a novel accuracy-driven optimization methodology that reduces HW resource requirements to support the deployment of HDC models in edge AI devices. In contrast to empirical approaches or exhaustive explorations, the proposed strategy employs a binary search of the hyper-parameters space and shows design runtime requirements that scale log-linearly with the workload complexity.

    \item \emph{MicroHD} concurrently co-optimizes multiple HDC hyper-parameters to achieve memory and performance improvements up to 266$\times$  when compared to standard HDC implementations. It is also agnostic of the workload characteristics and can hence be applied to HDC applications employing different encoding methods and input data.
    
    \item We also compare our optimized HDC models with results from previous works that employ binarization and dimensionality reduction to reduce resource requirements, showing up to 8$\times$ memory and computing requirements savings with respect to state-of-the-art (SoA) alternatives, for accuracy drops limited to 1\%. 
    
    \item Considering the latest trends in HDC developments, we show how \emph{MicroHD} can be beneficial for in-memory computing acceleration, by reducing area footprint and improving performance, and in federated learning settings, where our approach dramatically reduces the amount of data exchanged with the cloud.
\end{itemize}

The rest of the paper is organized as follows: first, we introduce the basics of HDC in Section~\ref{sec:background}. Then, we present related works in Section~\ref{sec:relatedworks}. Next, we introduce \emph{MicroHD} in Section~\ref{sec:method}. We discuss our conducted experiments in Section~\ref{sec:expsetup} and Section~\ref{sec:expresults} and finally draw our conclusions in Section~\ref{sec:conclusions}.

%% file: Background.tex
\section{Background on HD Computing}\label{sec:background}

Hyperdimensional computing (HDC) is a promising AI computing paradigm that enables efficient data processing and that stems from the neuroscience literature to emulate the functionalities of the human brain~\cite{kanerva2009hyperdimensional}. 
Several studies support the effectiveness of HDC algorithms in accomplishing common machine learning tasks such as text, image, and graph vertex classification~\cite{najafabadi2016hyperdimensional, 10024980, poduval2022graphd}, epileptic seizure detection~\cite{asgarinejad2020detection}, and human activity recognition~\cite{yao2021radar} among others. 
HDC performs distributed and usually low-precision data processing in a high-dimensional space where computation can be effectively parallelized using modern processing in-memory (PIM) accelerators~\cite{karunaratne2020memory}. 

The first stage of HDC requires input samples $x \in \mathcal{X}$ to be encoded using a specific function $\phi(x)$ which maps inputs $x$ from their generic $n$-dimensional space $\mathbb{R}^{n}$ to a much higher $d$-dimensional space $\mathcal{D} \subset \mathbb{R}^{d}$. A common practice is to define encoding functions that preserve similarity between input points $x$ and the corresponding encoded points $\phi(x) \in \mathcal{D}$. Mathematically, $\phi(a)\cdot\phi(b) \approx \mathcal{S}(a,b)$, where $\mathcal{S}$ is a desired similarity function. Usually, the elements of the hypervectors (HVs) may be constrained using binary (0, 1) or bipolar (-1, 1) representations to simplify integer arithmetic to bitwise logic operations, but at the cost of accuracy degradation~\cite{imani2019quanthd}.
Different encoding functions have been proposed in the literature, but all rely on some basic HDC operations: 

\begin{itemize}
    \item \textit{Binding}: $\otimes : \mathcal{D} \times \mathcal{D} \to \mathcal{D}$. Two input hypervectors are combined into a new hypervector \emph{dissimilar} to both inputs. For bipolar hypervectors, binding becomes an element-wise multiplication, while for binary hypervectors, it translates into a logical XOR.
    \item \textit{Bundling}: $\oplus : \mathcal{D} \times \mathcal{D} \to \mathcal{D}$. Two input hypervectors are combined into a new hypervector \emph{similar} to both inputs. Bundling is implemented through element-wise additions.
    \item \textit{Permutation}: $\rho : \mathcal{D} \to \mathcal{D}$. It implements a cyclic shift on the input hypervector and is usually employed to encode time-series data.
\end{itemize}

\begin{figure}[tp]
    \centering
    \includegraphics[width = 0.99\linewidth]{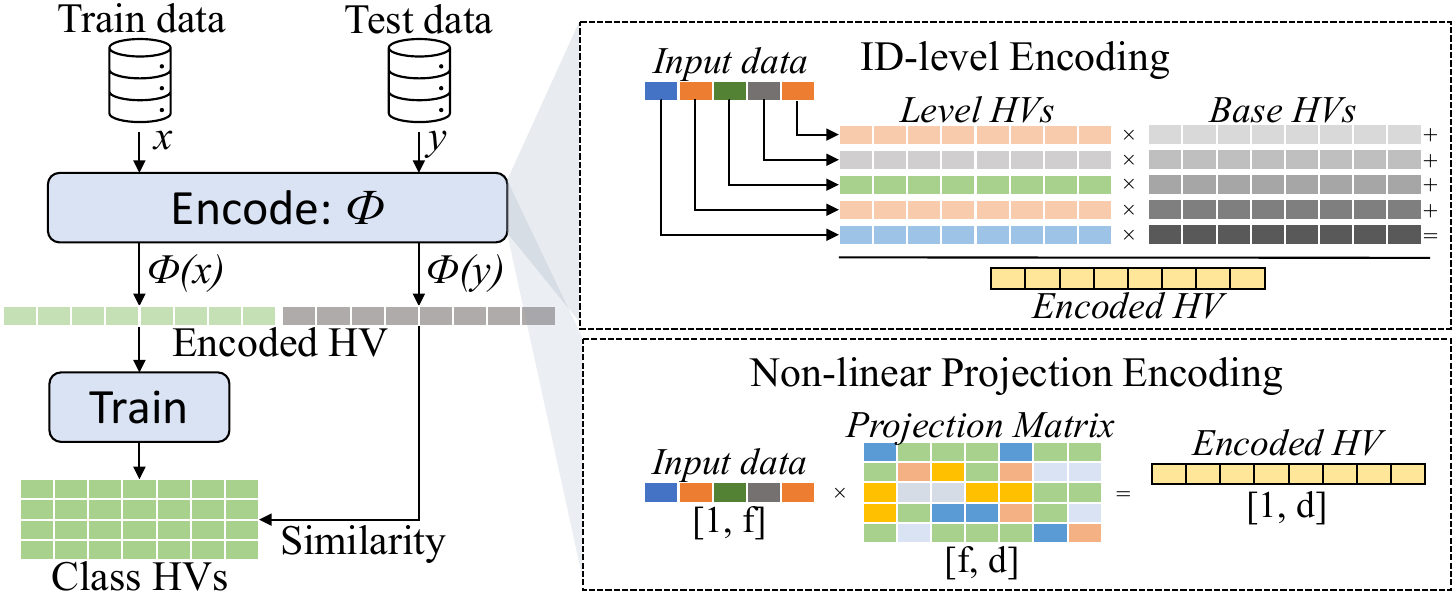}
    \caption{Basics of HDC (left). The encoding stage can be implemented using different $\phi$ functions, such as ID-level or projection methods (right).}
    \vspace{-0.1in}
    \label{fig:encoding}
    \vspace{-0.1in}
\end{figure}

Different encoding functions $\phi(x)$ have been proposed in the literature.
In ID-level encoding~\cite{rahimi2016hyperdimensional}, one random ID hypervector for each input feature is generated to act as a $d$-dimensional base in $\mathcal{D}$. Obtaining bases in $\mathcal{D}$ by random vector generation is possible because, in a sufficiently high dimensional space, two randomly generated vectors have the property to be (quasi-) orthogonal. Input feature values are then divided into $L$ levels. For each of them, a level HV is generated starting from a random HV $l_0$ for the lowest (or largest) feature value. Next, the other level HVs are obtained by iteratively flipping random $m$ bits from the previously generated level HV. This preserves a certain degree of similarity between two consecutive feature values encoded in $\mathcal{D}$, while ensuring the two extreme encoded level HVs are quite dissimilar.
Non-linear projection encoding only uses a matrix-vector multiplication to encode input values~\cite{thomas2021theoretical}. An input value $x$ is encoded in $\mathcal{D}$ by multiplying a matrix $\mathcal{P} \subset \mathbb{R}^{d \times n}$. While elements in $\mathcal{P}$ usually employ floating-point formats, using more compact integer or binary values has proven to reduce computing complexity at limited accuracy costs.
In Fig.~\ref{fig:encoding}, we provide an illustrative representation of the main flow in HDC training and inference (left), highlighting the two described encoding methods (right).

%% file: RelatedWorks.tex
\section{Related works on HDC optimization}\label{sec:relatedworks}

Several previous studies have noticed that traditional HDC implementations can be optimized for faster and more energy-efficient executions. From a hardware perspective, processing in-memory (PIM) accelerators can efficiently host HDC workloads and their use has become quite common in the last years. They can be designed using common SRAMs~\cite{rios2023bit}, or, more often, by employing new emerging memory technologies such as PCMs~\cite{karunaratne2020memory} and ReRAMs~\cite{dutta2022hdnn}, enabling high throughput, high parallelism, and low energy. 

Instead, from an application-level perspective, two main approaches have been explored to reduce the complexity of HDC workloads. Binarization quantizes encoded input and class HVs using 1-bit elements~\cite{hernandez2021onlinehd}. Such an approach significantly reduces memory requirements and transforms arithmetic operations such as additions and multiplications into much more efficient bitwise logic ones. Nevertheless, the impact on accuracy can be extremely high. To circumvent this issue, the authors of~\cite{imani2019quanthd} need to deploy both the floating-point and the binarized models to achieve high accuracy during on-device training, but this solution poses memory overheads that may be unacceptable for constrained TinyML systems. Similar observations are presented in~\cite{hernandez2021onlinehd} where the authors find that binarization implies the need for a higher dimensionality $d$ to limit accuracy degradation, whereas dimensionality can be effectively reduced only if employing an 8-bit (or higher) precision.
Finally, several works explore the effectiveness of dimensionality reductions to improve energy efficiency in different contexts. For example, the hybrid CNN-HDC accelerator presented in~\cite{dutta2022hdnn} runs HDC workloads employing an HD space of 4k dimensions. Similarly, the authors of~\cite{asgarinejad2020detection} focus on wearables for healthcare monitoring and limit the HD space to only 5k dimensions. A more extreme exploration of the accuracy/dimensionality pareto front is presented in~\cite{basaklar2021hypervector}, where the number of dimensions is reduced to the order of hundreds. Another very recent work~\cite{zeulin2023resource} proposes HDC for federated learning applications and shows latency and data transmission reductions when limiting $d$ to just 5k, 2.5k, 1k, or 0.5k.

\textit{Nevertheless, none of these works presents HDC optimizations driven by accuracy constraints}. As a consequence, output quality is not controlled and can exhibit drops often higher than 5\%, unacceptable for a multitude of different applications.

%% file: Methodology.tex
\section{HDC Optimization Methodology}\label{sec:method}

\begin{figure*}[tp]
    \centering
    \includegraphics[width = 0.99\linewidth]{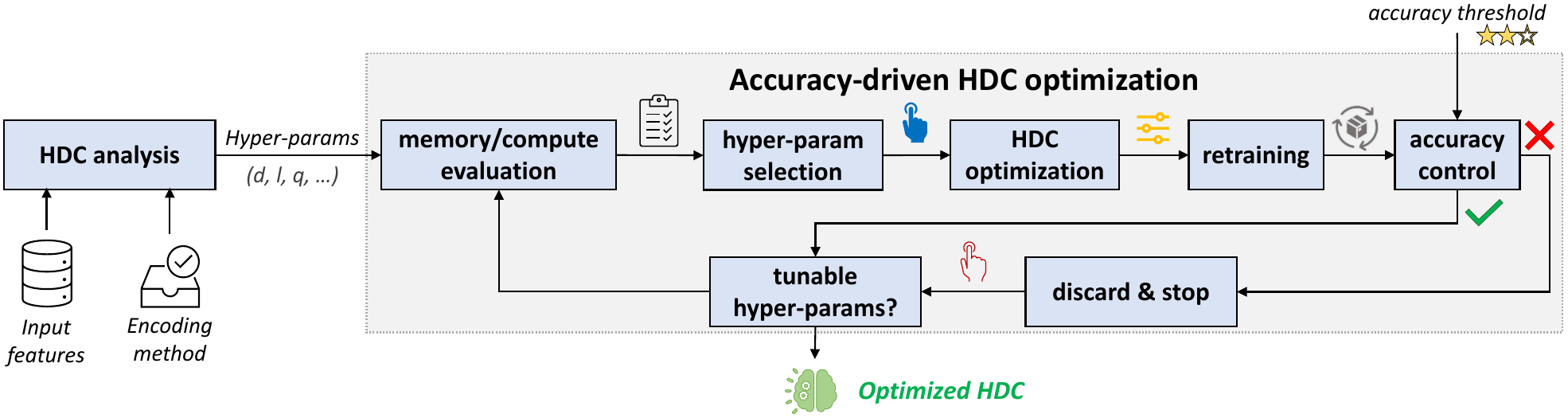}
    \vspace{-0.1in}
    \caption{\emph{MicroHD} optimization strategy. An initial analysis determines the impact of different HDC hyper-parameters on memory and computing requirements. Then, a greedy optimization step is applied and the model is re-trained. Finally, an accuracy check controls accuracy to user-defined levels and the procedure repeats until some tunable hyper-parameters exist.}
    \vspace{-0.1in}
    \label{fig:methodology}
\end{figure*}

\begin{table}[tp]
    \begin{center}
        \caption{Memory requirements (i.e., number of bits) of the two considered HDC encodings in terms of the number of input features $f$, dimensionality $d$, number of classes $c$, and quantization scheme $q$.}
        \label{tab:encoding_requirements}
        \vspace{-0.3cm}
        \begin{tabular}{lclc}
        \toprule
        \multicolumn{2}{c}{\textbf{ID-level}} & \multicolumn{2}{|c}{\textbf{Non-linear projection}} \\ \midrule
       
        \multicolumn{1}{l|}{ID HVs} & $ f \cdot d $ & \multicolumn{1}{|l|}{\multirow{2}{*}{$\mathcal{P} matrix$}} &  \multicolumn{1}{c}{\multirow{2}{*}{$ f \cdot d \cdot q$}}       \\
        \multicolumn{1}{l|}{Level HVs} & $ l \cdot d $ & \multicolumn{1}{|c|}{} & \multicolumn{1}{c}{}        \\
        \multicolumn{1}{l|}{Class HVs} & $ c \cdot d \cdot q $ & \multicolumn{1}{|l|}{Class HVs}  & $ c \cdot d \cdot q $     \\
        \multicolumn{1}{l|}{Total} & $ d \cdot (f + l + cq) $ & \multicolumn{1}{|l|}{Total} & $d \cdot q \cdot (f + c)$    \\
        \bottomrule
        \end{tabular}
        \vspace{-0.1in}
    \end{center}
\end{table}

\subsection{Resource requirements analysis}
The amount of memory resources and computation HDC requires in encoding, training, and inference depends on multiple variables, with only some of them defined when designing the model. For example, the number of features in the original input space determines the number of base vectors in the encoding process and is usually a constant that cannot be optimized. Nevertheless, encoding methods and the resulting HDC model have the largest impact on resource requirements and their design is key to achieving higher energy efficiency levels. 
Therefore, we herein analyze two common encoding methods (i.e., ID-level and non-linear projection) in terms of resource requirements and summarize them in Table~\ref{tab:encoding_requirements}.

\subsubsection{ID-level encoding}
In ID-level encoding, memory and computing requirements depend on the number of input features $f$, the hyperspace dimensionality $d$, the number of generated level hypervectors $l$, and the quantization level $q$. While $f$ is mainly related to the target application and may not be adjusted in general, HDC performance can be optimized by tuning the other hyper-parameters. Here, we assume base and level hypervectors, as well as encoded input samples, employ a bipolar representation, hence only requiring one bit for each dimension (i.e., $q=1$). Instead, class hyper-vectors, generated via the accumulation of encoded input data, are assumed to have a compact 16-bit integer representation to maximize accuracy. It can be noted from Table~\ref{tab:encoding_requirements} that $d$ has the largest impact on resource requirements.


\subsubsection{Non-linear projection encoding}
In projection encoding, base and level hypervectors are replaced by a projection matrix $\mathcal{P}$, with the encoding phase being a matrix-vector multiplication between $\mathcal{P}$ and the input sample $x$. In general, $\mathcal{P}$ contains floating-point values, but, since quantization is commonly employed, we consider also in this case 16-bit integer representations as a baseline for elements in $\mathcal{P}$. Therefore, memory and computing requirements mainly depend on $d$ and $f$, which are two dimensions of the projection matrix $\mathcal{P}$, and on the employed quantization bitwidth $q$.
\\

In both cases, our methodology also evaluates computing requirements considering (i) the encoding of an individual input sample, (ii) its comparison with class hypervectors to evaluate similarity scores (i.e., inference), and (iii) its addition to class HVs to update the current model (i.e., single-pass training). Since base and level hypervectors are bipolar, while class hypervectors (and the $\mathcal{P}$ matrix) use more precise integer formats, we define the computing requirements of the two encoding methods in terms of the number of operations (i.e., binding and bundling) per bit and use them as a proxy for performance.

\subsection{Optimization approach}
An overview of the proposed accuracy-driven optimization methodology is depicted in Fig~\ref{fig:methodology}. As an initial step, \emph{MicroHD} analyzes the input workload to determine which hyper-parameters can be tuned to reduce memory and computing requirements. Then, the selected hyper-parameters are used as inputs to the main optimization loop that iteratively reduces HDC complexity. First, memory and computing requirements of the current HDC model are computed. Then, a greedy-based selector determines the hyper-parameter whose optimization in the current iteration would lead to the largest efficiency improvements. Consequently, this optimization step is applied to the current model, which is then retrained for a limited number of epochs to regain accuracy. Finally, a user-defined accuracy constraint is used to determine whether the obtained model satisfies the desired quality of service (QoS) or not. In the former case, the obtained model is used as input for the next iteration and new hyper-parameters are evaluated to optimize the model. On the opposite, the applied optimization step is discarded. The optimization loop terminates when no HDC hyper-parameter can be optimized more.

To reduce the run-time of \emph{MicroHD} and to enable large-scale HDC workload optimizations, we implement a binary search of the optimal values for each considered hyper-parameter. To this end, we list a number of admitted values $V$ for each hyper-parameter in ascending order, with the last element being the baseline value for the corresponding hyper-parameter (e.g., 10k, for the hyperspace dimensionality). Then, we implement the discussed optimization approach, where, at each iteration, the tested value for the target hyper-parameter is selected by using a binary search.
When a certain hyper-parameter is successfully optimized in the previous iteration, the binary search looks for smaller values (i.e., moving to the left of the list), while it looks for larger values in the opposite case (i.e., moving to the right).

Finally, in addition to the accuracy control, another novel aspect of \emph{MicroHD} is the co-optimization of different hyper-parameters. In fact, for each of them, the target value to be used to optimize the HDC workload is evaluated in terms of resource requirements savings at the beginning of each iteration. Then, following a greedy approach, the one that could lead to the largest gains is selected and used to optimize the model. Our results demonstrate that this approach leads to improvements larger than the current state-of-the-art (see Section~\ref{sec:comparison_soa}).

\subsection{Complexity analysis}
We herein discuss the complexity of the proposed methodology from an algorithmic perspective. 
The two key elements of \emph{MicroHD} are the binary search of the hyper-parameters space and the greedy optimization step. Both alleviate the design complexity of the proposed approach and make it a better (and practical) alternative to exhaustive explorations. On one hand, full explorations can provide optimal hyper-parameter configurations. On the other hand, these solutions are not useful in practice, as complexity increases exponentially with the number of considered hyper-parameters. On the opposite, \emph{MicroHD} scales only log-linearly in the number of considered hyper-parameters, thus allowing the optimization of large HDC workloads at acceptable times. Assuming $H$ hyper-parameters, each of them allowing the same number of different values $V$, the complexity of an exhaustive search would be $V^H$. Conversely, \emph{MicroHD} explores the hyper-parameters space with a complexity in the order of $H\log_{2}V$, thus allowing the optimization of large HDC workloads in acceptable times. Notice that, since \emph{MicroHD} is run offline and serves as an optimization tool for baseline HDC implementations, the required runtime of its binary search of the hyper-parameter space does not represent an overhead to the optimized HDC model online performance.

%% file: ExpSetup.tex
\section{Experimental Setup}\label{sec:expsetup}

The proposed methodology is evaluated considering ID-level~\cite{rahimi2016hyperdimensional} and non-linear projection~\cite{thomas2021theoretical} encodings on the following datasets: ISOLET~\cite{asuncion2007uci}, UCIHAR~\cite{anguita2013public}, MNIST~\cite{deng2012mnist}, FMNIST~\cite{xiao2017fashion}, PAMAP~\cite{reiss2012introducing}, and Connect-4~\cite{asuncion2007uci}. The training set of each dataset is split into training and evaluation data using an 80/20 split ratio: training data is used to train the optimized HDC models, while evaluation data is used to estimate accuracy as part of the MicroHD methodology. Reported accuracy results are obtained over the testing set.
The evaluation framework is developed in PyTorch, including functionalities from the TorchHD library~\cite{heddes2023torchhd} for dataset loaders and encoding methods. We employ a retraining strategy based on the approach proposed in~\cite{hernandez2021onlinehd}. We set a fixed learning rate $lr=1$ and a number of retraining epochs $ep=30$, in line with what previous works proposed. Memory requirements are computed considering both the size of the HDC model (i.e., the class HVs) and the HVs required for the encoding stage, while computing complexity is evaluated considering the number of binding and bundling operations per bit, as discussed in Section~\ref{sec:method}.

Experiments are run on an NVIDIA GeForce RTX 4090 GPU equipped with 24GB of memory and all the optimized models presented in this work can be obtained in less than one day. To highlight the performance gain in both low-power and high-performance hardware, we perform a runtime analysis on the following hardware: NVIDIA GeForce RTX 4090 GPU, Intel i7, ARM-Cortex A7, and ARM-Cortex M4.
Baseline hyper-parameter configurations are extracted from previous works and assume a dimensionality $d=10000$, a number of distinct level hypervectors $l=1024$, and a quantization bitwidth $q=16$ for class hypervectors and for the non-linear projection matrix $\mathcal{P}$. We observe no accuracy improvements when increasing any of these hyper-parameters. Finally, we validate our proposed approach constraining accuracy using three different thresholds of 0.5\%, 1.0\%, and 5\%. 
Experiments comparing our results with previous works consider the best SoA HDC implementations with optimized hyper-parameters and showing accuracy degradations close to the ones investigated in this work.

%% file: ExpResults.tex
\section{Experimental Results}\label{sec:expresults}

\begin{figure}[tp]
    \centering
    \includegraphics[width = 0.99\linewidth]{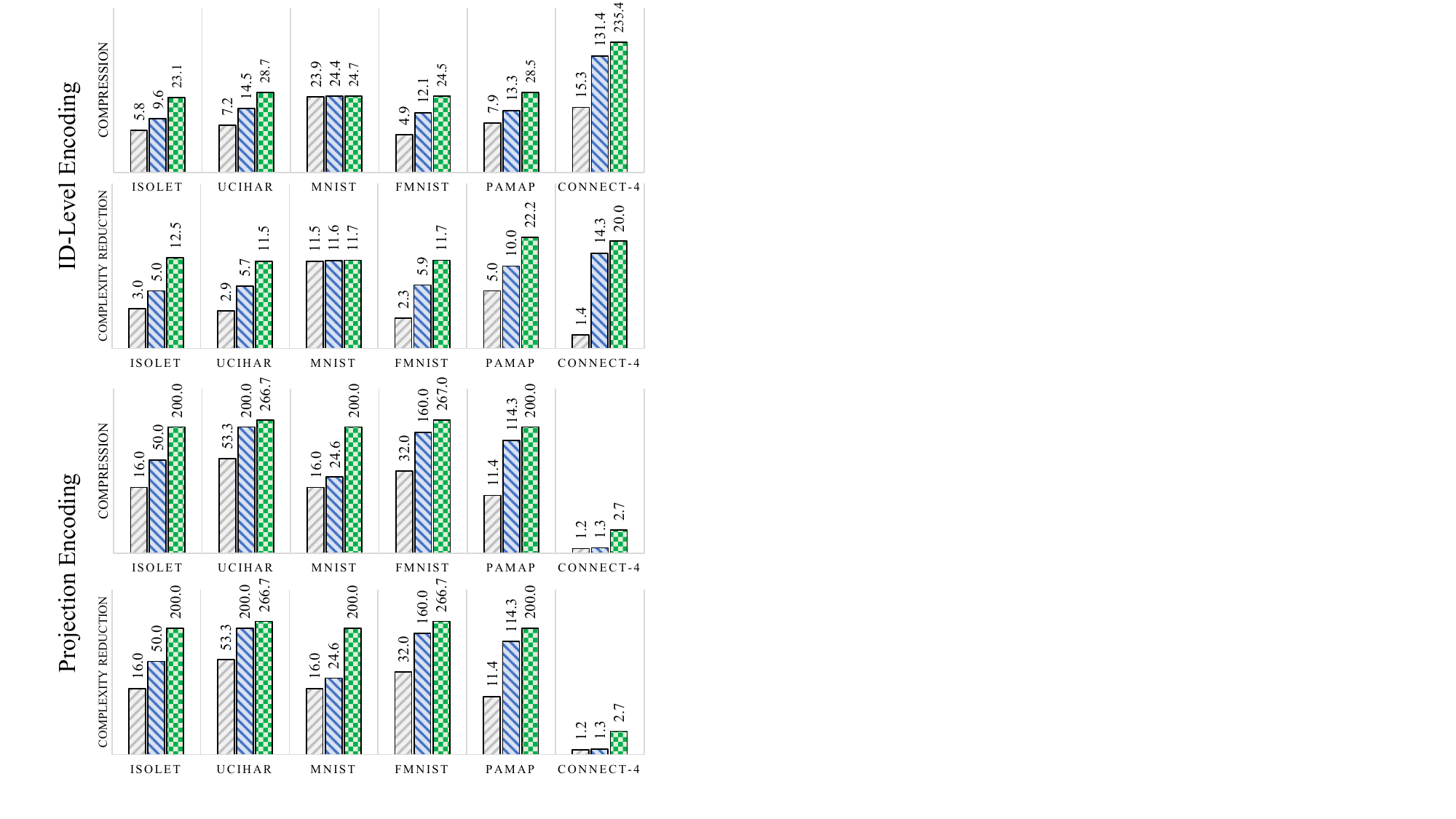}
    \vspace{-0.1in}
    \caption{Achieved compression and workload reduction factors in HDC models using ID-Level and non-linear projection encoding evaluated on multiple datasets imposing a 0.5\% (grey), 1.0\% (blue), and 5.0\% (green) accuracy threshold. }
    \label{fig:results_compression}
    \vspace{-0.2in}
\end{figure}


\subsection{Compression and runtime improvements}

The achieved improvements in terms of memory and computing requirements are depicted in Fig.~\ref{fig:results_compression}. For each dataset, the colored bars in each plot refer to the three considered accuracy thresholds of 0.5\%, 1.0\%, and 5.0\%. These results demonstrate that \emph{MicroHD} effectively improves baseline HDC, achieving up to 235$\times$/22$\times$ and 267$\times$/267$\times$ reductions in memory/computing requirements in HDC workloads employing ID-level and non-linear encodings, respectively. In particular, we observe improvements of up to 53$\times$ even for a very conservative accuracy threshold of 0.5\%.

\begin{figure}[tp]
    \centering
    \includegraphics[width = 0.99\linewidth]{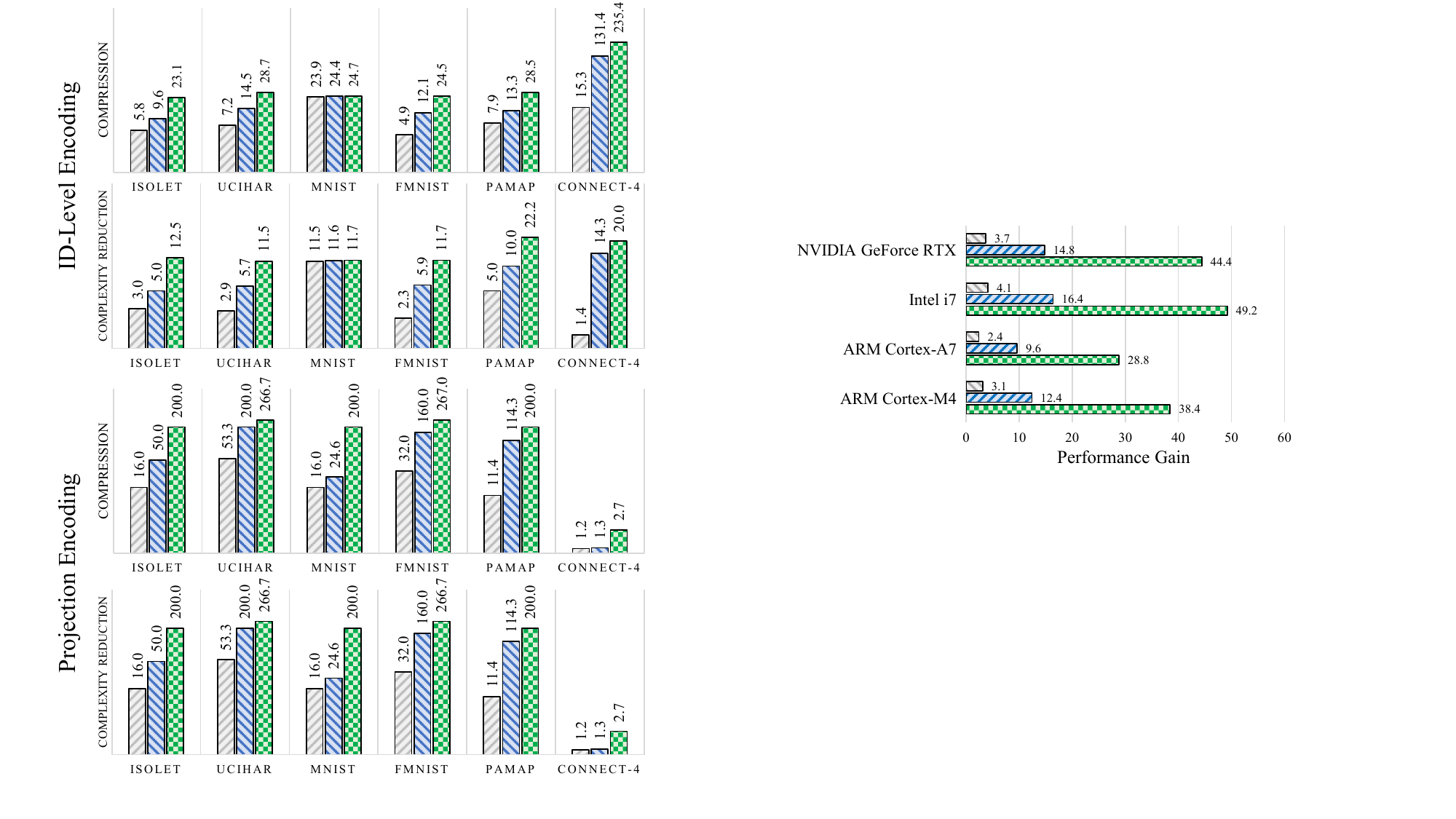}
    \vspace{-0.1in}
    \caption{MicroHD optimized average performance gain factor over the considered benchmarks for accuracy thresholds of 0.5\% (grey), 1.0\% (blue), and 5.0\% (green).}
    \label{fig:performance}
    \vspace{-0.2in}
\end{figure}

These findings confirm that baseline HDC models are often highly oversized for most workloads, thus leaving the door open for significant resource optimizations.
Nevertheless, we also observe how this is not the case in general, underlying how critical is an accuracy-driven approach to maintaining a desired degree of QoS. In particular, our results indicate that benchmark complexity and encoding method play an important role in determining the pareto front in the accuracy/resource trade-off. For example, in contrast to average improvements obtained in other benchmarks, only a 1.3$\times$ gain is achieved for a 1\% accuracy drop on the Connect-4 dataset when employing a non-linear encoding. However, this anomaly is not observed when employing an ID-level encoding where, on the opposite, the largest gains (i.e., 131$\times$) are retrieved.

A more detailed overview of the optimized HDC models including accuracy, optimized hyper-parameters, and resulting memory occupation is reported in Table~\ref{tab:results_params}, considering experiments using an accuracy constraint of 1\%. We observe that for HDC workloads employing an ID-level encoding, the hyperspace dimensionality can be reduced to just 1k or 2k dimensions in all cases. However, the obtained number of level HVs and the quantization levels largely differ across the considered benchmarks, suggesting once again that different HDC workloads exhibit different degrees of robustness against the optimization of HDC hyper-parameters. Moreover, dimensionality is usually the metric more often targeted in initial iterations due to its higher impact on memory and computing requirements (see Table~\ref{tab:encoding_requirements}).
Overall, the obtained HDC models require from as few as 10KB to just a few hundred KB of memory for their end-to-end execution, demonstrating how \emph{MicroHD} effectively supports energy-efficient edge AI deployment.

Performance evaluations on different hardware resources are summarized in Figure~\ref{fig:performance}, consider the analyzed encoding methods and datasets, and reveal average improvements of 3.3$\times$, 13.3$\times$, and 40.2$\times$, for accuracy loss limited to 0.5\%, 1.0\%, and 5.0\%, respectively. These improvements are mainly due to dimensionality reduction because quantization, especially for few-bit quantization schemes (e.g., 2 bits or 3 bits), cannot be fully leveraged in MCUs or GPUs.

\begin{table*}[tp]
    \begin{center}
        \caption{Accuracy, hyper-parameters, and memory occupation of our optimized benchmarks employing ID-level or non-linear projection encoding methods, obtained imposing a 1\% accuracy threshold.}
        \label{tab:results_params}
        \vspace{-0.4cm}
        \begin{tabular}{lccccccccccccc}
        \toprule
        \multicolumn{1}{c|}{} & \multicolumn{7}{c}{\textbf{ID-level encoding}} & \multicolumn{6}{|c}{\textbf{Non-linear projection encoding}} \\ \midrule

        \multicolumn{1}{c|}{\multirow{2}{*}{\textbf{Dataset}}} & \multicolumn{2}{c}{Accuracy (\%)}  & \multicolumn{3}{c}{\textit{Hyper-parameters}} &  \multicolumn{2}{c}{Memory Size [KB]} & \multicolumn{2}{|c}{Accuracy (\%)} & \multicolumn{2}{c}{\textit{Hyper-parameters}} & \multicolumn{2}{c}{Memory size [KB]} \\ 

        \multicolumn{1}{c|}{} & \textit{base} & \textit{MicroHD} & \textit{d} & \textit{l} & \textit{q} & \textit{base} & \textit{MicroHD} & \multicolumn{1}{|c}{\textit{base}} & \textit{MicroHD} & \textit{d} & \textit{q} & \textit{base} & \textit{MicroHD} \\
        
        \midrule

        \multicolumn{1}{c|}{ISOLET} & 91.41 & 90.45 & 2000 & 32 & 16 & 2530.5 & 263.9 & \multicolumn{1}{|c}{93.39} & 92.51 & 200 & 16  & 12578.1 & 251.5      \\
        \multicolumn{1}{c|}{UCIHAR} & 90.40 & 89.48 & 2000  & 4 & 3 & 2071.5 & 143.0 & \multicolumn{1}{|c}{91.31} & 90.33 & 200 & 4   & 11093.7 & 55.5     \\
        \multicolumn{1}{c|}{MNIST} & 86.77 & 85.83 & 1000  & 4 & 2 & 2421.9 & 99.4 & \multicolumn{1}{|c}{92.50} & 91.57 & 500 & 13  & 15527.3 & 630.8       \\
        \multicolumn{1}{c|}{FMNIST} & 79.62 & 78.92 & 2000  & 16  & 2 & 2421.9 & 200.7 & \multicolumn{1}{|c}{78.56} & 77.76 & 200 & 5   & 15527.3 & 97.1     \\
        \multicolumn{1}{c|}{PAMAP} & 91.47 & 90.57 & 1000 & 512 & 16 & 2530.5 & 190.5 & \multicolumn{1}{|c}{92.65} & 92.05 & 200 & 7   & 1406.3 & 12.3      \\
        \multicolumn{1}{c|}{Connect-4} & 76.71 & 75.87 & 1000  & 16 & 7 & 1379.4 & 10.5 & \multicolumn{1}{|c}{89.92} & 89.07 & 9000 & 14 & 898.4 & 707.5        \\
        \bottomrule
        \end{tabular}
    \end{center}
    \vspace{-0.3cm}
\end{table*}

\subsubsection{Impact on PIM-based HDC accelerators}
PIM accelerators for HDC workloads greatly benefit from the proposed methodology. Indeed, state-of-the-art accelerators based on RRAMs or PCMs usually consider HDC workloads in a 10k-dimensional space~\cite{karunaratne2020memory, schindler2021primer}. Since deploying such a high number of bitcells allows parallel computation across all HDC dimensions, performance may not be significantly improved in our HDC models\footnote{Assuming single-cycle computation for all dimensions.}. On the contrary, the much lower memory requirements of our optimized benchmarks support large area footprint reductions that also lead to energy savings, critical metrics for TinyML systems. Both improvements are proportional to the compression results presented in this section, with the number of bitcells being reduced by the same discussed factors (e.g., up to 266$\times$), while the overall area footprint reduction may differ according to specific PIM implementations, as different designs introduce custom peripheral logic for HDC operations. Nevertheless, in recent implementations peripheral logic is extremely lightweight, with the total area being dominated by the size of deployed memory arrays that account for more than 97\% of the overall area footprint~\cite{khaleghi2021tiny}.

\subsubsection{Impact on FL-based TinyML}
\emph{MicroHD} can be also employed to improve latency and energy in federated learning (FL) configurations. 
We consider~\cite{zeulin2023resource} as a baseline implementation to evaluate the benefits of our strategy in this scenario. This very recent work optimizes communication rounds by collaboratively training $M$ independent HDC sub-models, ultimately limiting the amount of transferred data when compared to standard HDC federated learning settings~\cite{zhang2023hyperdimensional}. It reduces the hyperspace dimensionality to as few as 0.5k dimensions, but integer values are required to avoid critical accuracy degradations. When using \emph{MicroHD} to optimize their benchmarks on top of their proposed FL settings, we achieve an average 3.3$\times$ lower data communication with the cloud, for similar accuracy levels\footnote{Analysis over the common dataset and considering non-linear encoding HDC with $d$=1k as the best implementations in~\cite{zeulin2023resource}}.

\subsection{Comparison with previous works}\label{sec:comparison_soa}

As previously discussed, previous works do not control accuracy when applying dimensionality reduction or quantization. In those cases, only impractical exhaustive design-space exploration could provide a pareto front from which the best candidate under desired accuracy constraints can be determined. This limitation also affects the exploration depth performed in previous works, where only a few configurations can be evaluated. As a consequence, we cannot perform a fair comparison by measuring the obtained gains for target output accuracy levels. Therefore, we compare our results with previous works that perform HDC optimizations, and, from those works, we consider the best HDC model configurations showing accuracy degradations close to 1\%. A summary of this comparative analysis showing the resulting memory requirements and accuracy degradation is presented in Table~\ref{tab:soa_comparison}.

Our results outperform previous best implementations by achieving average memory requirements reductions of 3.9$\times$ across the evaluated benchmarks. This finding shows that the proposed greedy hyper-parameters search can be highly effective in achieving model compression thanks to its holistic optimization approach which concurrently evaluates and optimizes multiple hyper-parameters. Moreover, \emph{MicroHD} also offers the fastest design runtime while being the only one that ensures target accuracy levels.

\begin{table}[tp]
    \begin{center}
        \caption{Comparing our optimized HDC models with previous works in terms of memory requirements (KB) and accuracy drop (\%) compared to baseline top-accuracy implementations.}
        \label{tab:soa_comparison}
        \vspace{-0.3cm}
        \begin{tabular}{lccccc}
        \toprule
                
        \multicolumn{1}{l}{\multirow{2}{*}{\textbf{Work}}} & \multicolumn{1}{c}{ISOLET}  & \multicolumn{1}{c}{UCIHAR} &  \multicolumn{1}{c}{MNIST} & \multicolumn{1}{c}{FMNIST} & \multicolumn{1}{c}{PAMAP} \\

        \multicolumn{1}{c}{} & \multicolumn{5}{c}{\textit{memory size [KB] / accuracy drop (\%)}} \\
        
        
        \midrule

        \multicolumn{1}{l}{\cite{hernandez2021onlinehd}} & 471/1 & - & - & - & -   \\
        \multicolumn{1}{l}{\cite{imani2019quanthd}} & 628/2 & 693/3 & -  & - & 88/3  \\
        \multicolumn{1}{l}{\cite{basaklar2021hypervector}} & - & 71/1 & - & - & -   \\
        \multicolumn{1}{l}{\cite{zeulin2023resource}} & - & 554/6 & 786/1 & 776/1 & - \\
        \multicolumn{1}{l}{\emph{MicroHD}} & \textbf{251/1} & \textbf{55/1} & \textbf{631/1} & \textbf{97/1} & \textbf{12/1} \\
        \bottomrule
        \end{tabular}
        \vspace{-0.2in}
    \end{center}
\end{table}

%% file: Conclusions.tex
\section{Conclusions}\label{sec:conclusions}

In this work, we presented \emph{MicroHD}, a novel optimization methodology for HDC workloads. \emph{MicroHD} is the first approach that directly controls accuracy degradation by implementing an iterative optimization procedure. The tuning of HDC hyper-parameters combines a binary search over the target hyperdimensional space with a greedy optimization step, enabling good performance and scalability. Compared to baseline HDC models, \emph{MicroHD} obtains up to 250$\times$ higher performance and 266$\times$ lower memory requirements, hence supporting the real-life deployment of HDC-based TinyML applications. \emph{MicroHD} co-optimizes HDC hyper-parameters to achieve top improvements in terms of memory and computing requirements. Comparisons with the state-of-the-art demonstrate that our optimized models achieve 3.3$\times$ higher compressions, on average, on common evaluated benchmarks. Finally, we studied the benefits of our approach for state-of-the-art PIM acceleration, reducing the number of required bitcells by up to 266$\times$, and in emerging HDC-based FL settings, enabling up to 3.3$\times$ lower energy cost and latency for communication rounds with the cloud.

%% file: acknowledgements.tex
\section{Acknowledgements}\label{sec:ack}
This work was supported in part by National Science Foundation under Grants \#2112665 (TILOS AI Research Institute), \#2003279, \#1911095, \#1826967, \#2100237, \#2112167, and in part by PRISM and CoCoSys, centers in JUMP 2.0, an SRC program sponsored by DARPA.